\documentclass{article}%
\usepackage{amsmath}
\usepackage{amsfonts}
\usepackage{amssymb}
\usepackage{graphicx}%
\providecommand{\U}[1]{\protect\rule{.1in}{.1in}}

\begin{document}

\author{Giuseppe Castagnoli
\and Pieve Ligure, Italy, giuseppe.castagnoli@gmail.com}
\title{Discussing the explanation of the quantum speed up}
\maketitle

\begin{abstract}
In former work, we showed that a quantum algorithm is the sum over the
histories of a classical algorithm that knows in advance 50\% of the
information about the solution of the problem -- each history is a possible
way of getting the advanced information and a possible result of computing the
missing information. We gave a theoretical justification of this 50\% advanced
information rule and checked that it holds for a large variety of quantum
algorithms. Now we discuss the theoretical justification in further detail and
counter a possible objection. We show that the rule is the generalization of a
simple, well known, explanation of quantum nonlocality -- where logical
correlation between measurement outcomes is physically backed by a
causal/deterministic/local process with causality allowed to go backward in
time with backdated state vector reduction. The possible objection is that
quantum algorithms often produce the solution of the problem in an apparently
deterministic way (when their unitary part produces an eigenstate of the
observable to be measured and measurement produces the corresponding
eigenvalue \ -- the solution -- with probability 1), while the present
explanation of the speed up relies on the nondeterministic character of
quantum measurement. We show that this objection would mistake the
nondeterministic production of a definite outcome for a deterministic production.

\end{abstract}

The "50\% advanced information rule" formulated in $\left[  1\right]  $ and
$\left[  2\right]  $ says that a quantum algorithm can be broken down into a
sum over the histories of a classical algorithm that knows in advance 50\% of
the information about the solution of the problem. Each history is a possible
way of getting the advanced information and a possible result of computing the
missing information. This rule explains the quantum speed up, the fact that
quantum algorithms require a lower number of operations than their classical
counterparts. We gave a theoretical justification of the rule and checked that
the rule holds for a large variety of quantum algorithms. In the following, we
review in further detail the theoretical justification, focusing on Grover's
data base search algorithm.

First we review Grover's algorithm in the simple instance of database size
$N=4$. The exposition should be such that no previous knowledge of quantum
computer science is required.

Thus, we have a problem and the algorithm that solves the problem. The problem
is defined as follows -- we resort to a visualization to aid intuition. We
have a chest of 4 drawers numbered 00, 01, 10, 11, a ball, and two players.
The first player (the oracle) hides the ball in drawer number $\mathbf{k}%
\equiv\mathbf{~}k_{0},k_{1}$, chosen at random, and gives to the second player
the chest of drawers. This is represented by a black box that, given an input
$\mathbf{x}\equiv x_{0},x_{1}$ (a drawer number), computes the Kronecker
function $\delta\left(  \mathbf{k},\mathbf{x}\right)  $ (1 if $\mathbf{k}%
=\mathbf{x}$, 0 otherwise). The second player -- the algorithm -- should find
the drawer with the ball, i.e. specify its number, and this is done by
computing $\delta\left(  \mathbf{k},\mathbf{x}\right)  $ for different values
of $\mathbf{x}$ -- by opening different drawers. A classical algorithm
requires 2.25 computations of $\delta\left(  \mathbf{k},\mathbf{x}\right)
$\ on average -- 3 computations if one wants to be a priori certain of finding
the solution. The quantum algorithm yields the solution with certainty with
just 1 computation -- see $\left[  3\right]  $.

In our representation of the quantum algorithm, the quantum computer has three
registers. A two qubit register $K$ contains the oracle's choice of the value
of $\mathbf{k}$, the first input of the computation of $\delta\left(
\mathbf{k},\mathbf{x}\right)  $. The state of this register can be $\left\vert
00\right\rangle _{K}$, or $\left\vert 01\right\rangle _{K}$, etc., which means
oracle's choice $\mathbf{k}=00$,\ or $\mathbf{k}=01$, etc.; of course, we can
also have a superposition of such sharp quantum states. A two qubit register
$X$ contains the argument $\mathbf{x}$ to query the black box with -- the
other input of the computation of $\delta\left(  \mathbf{k},\mathbf{x}\right)
$. A one qubit register $V$ is meant to contain the result of the computation,
modulo 2 added to its initial content for logical reversibility. The three
registers undergo a suitable unitary evolution, where in particular
$\delta\left(  \mathbf{k},\mathbf{x}\right)  $ is computed once. Measuring the
content of register $K$ yields the oracle's choice $\mathbf{k}$; this
measurement can be performed, indifferently, at the beginning or at the end of
the algorithm. Measuring the content of register $X$ at the end of the
algorithm yields the solution of the problem $\mathbf{x}=\mathbf{k}$.

The initial state of the computer registers is:%

\begin{equation}
\frac{1}{4\sqrt{2}}\left(  \left\vert 00\right\rangle _{K}+\left\vert
01\right\rangle _{K}+\left\vert 10\right\rangle _{K}+\left\vert
11\right\rangle _{K}\right)  \left(  \left\vert 00\right\rangle _{X}%
+\left\vert 01\right\rangle _{X}+\left\vert 10\right\rangle _{X}+\left\vert
11\right\rangle _{X}\right)  \left(  \left\vert 0\right\rangle _{V}-\left\vert
1\right\rangle _{V}\right)  , \label{input}%
\end{equation}
all registers are prepared in even weighted superpositions of all
possibilities. This state is the input of the computation of $\delta\left(
\mathbf{k},\mathbf{x}\right)  $. This means that the computation will be
performed in quantum parallelism on each and every term of the superposition.
Let us consider for example the input term $-\left\vert 01\right\rangle
_{K}\left\vert 01\right\rangle _{X}\left\vert 1\right\rangle _{V}$. It means
that the input of the black box is $\mathbf{k}=01$\ and $\mathbf{x}=01$ and
that the initial content of register $V$\ is 1. The computation yields
$\delta\left(  01,01\right)  =1$, which modulo 2 added to the initial content
of $V$\ yields the output term $-\left\vert 01\right\rangle _{K}\left\vert
01\right\rangle _{X}\left\vert 0\right\rangle _{V}$ (registers $K$\ and $X$
keep the memory of the input, for logical reversibility). Similarly the input
term $\left\vert 01\right\rangle _{K}\left\vert 01\right\rangle _{X}\left\vert
0\right\rangle _{V}$ is transformed into the output term $\left\vert
01\right\rangle _{K}\left\vert 01\right\rangle _{X}\left\vert 1\right\rangle
_{V}$. Summing up, $\left\vert 01\right\rangle _{K}\left\vert 01\right\rangle
_{X}\left(  \left\vert 0\right\rangle _{V}-\left\vert 1\right\rangle
_{V}\right)  $ is transformed into $-\left\vert 01\right\rangle _{K}\left\vert
01\right\rangle _{X}\left(  \left\vert 0\right\rangle _{V}-\left\vert
1\right\rangle _{V}\right)  $. The computation of $\delta\left(
\mathbf{k},\mathbf{x}\right)  $\ inverts the phase of those terms where
$\mathbf{k}=\mathbf{x}$, leaving the other terms unaltered. In the overall, it
changes (\ref{input}) into:%

\begin{equation}
\frac{1}{4\sqrt{2}}\left[
\begin{array}
[c]{c}%
\left\vert 00\right\rangle _{K}\left(  -\left\vert 00\right\rangle
_{X}+\left\vert 01\right\rangle _{X}+\left\vert 10\right\rangle _{X}%
+\left\vert 11\right\rangle _{X}\right)  +\\
\left\vert 01\right\rangle _{K}\left(  \left\vert 00\right\rangle
_{X}-\left\vert 01\right\rangle _{X}+\left\vert 10\right\rangle _{X}%
+\left\vert 11\right\rangle _{X}\right)  +\\
\left\vert 10\right\rangle _{K}\left(  \left\vert 00\right\rangle
_{X}+\left\vert 01\right\rangle _{X}-\left\vert 10\right\rangle _{X}%
+\left\vert 11\right\rangle _{X}\right)  +\\
\left\vert 11\right\rangle _{K}\left(  \left\vert 00\right\rangle
_{X}+\left\vert 01\right\rangle _{X}+\left\vert 10\right\rangle _{X}%
-\left\vert 11\right\rangle _{X}\right)
\end{array}
\right]  \left(  \left\vert 0\right\rangle _{V}-\left\vert 1\right\rangle
_{V}\right)  , \label{secondstage}%
\end{equation}
where four orthogonal states of $K$\ , each corresponding to a single value of
$\mathbf{k}$, are correlated with four orthogonal states of $X$. This means
that the information about the value of $\mathbf{k}$\ has propagated to
register $X$.

A suitable rotation of the\ measurement basis of $X$ transforms entanglement
between registers $K$\ and $X$\ into correlation between the outcomes of
measuring their contents, transforming (\ref{secondstage}) into:%

\begin{equation}
\frac{1}{2\sqrt{2}}\left(  \left\vert 00\right\rangle _{K}\left\vert
00\right\rangle _{X}+\left\vert 01\right\rangle _{K}\left\vert 01\right\rangle
_{X}+\left\vert 10\right\rangle _{K}\left\vert 10\right\rangle _{X}+\left\vert
11\right\rangle _{K}\left\vert 11\right\rangle _{X}\right)  \left(  \left\vert
0\right\rangle _{V}-\left\vert 1\right\rangle _{V}\right)  \label{output}%
\end{equation}

The solution is in register $X$. We incidentally note that the unitary
transformation of (\ref{input}) into (\ref{output}) is the identity in the
Hilbert space of register $K$.

The oracle's choice has not been performed as yet. It is performed by
measuring $\left[  K\right]  $, the content of register $K$, in (\ref{output})
or, indifferently, (\ref{input}). Say that we obtain $\ \mathbf{k}=01$. State
(\ref{output}) reduces to%

\begin{equation}
\frac{1}{\sqrt{2}}\left\vert 01\right\rangle _{K}\left\vert 01\right\rangle
_{X}\left(  \left\vert 0\right\rangle _{V}-\left\vert 1\right\rangle
_{V}\right)  . \label{am}%
\end{equation}

Measuring $\left[  X\right]  $\ in (\ref{am}) yields the solution produced by
the algorithm, namely the eigenvalue $\mathbf{x}$ $=01$. We can say that the
oracle's choice of the drawer number 01 implies that the algorithm outputs 01.
However, instead of measuring $\left[  K\right]  $ in (\ref{output}), we could
have measured $\left[  X\right]  $, obtaining, say, $\mathbf{x}$ $=01$, which
means state reduction on (\ref{am}) again. Measuring $\left[  K\right]  $ in
(\ref{am})\ yields $\mathbf{k}=01$. In this case we can say that reading the
output of the algorithm and finding 01 implies that the oracle's choice is 01.
In fact there is mutual implication between the two measurement outcomes. In
the following we discuss the relationship between the logical notion of
implication and the physical notion of causality, arguing that there must be a
causal/deterministic/local process that physically backs logical implication.
We will see that there is always such a process, provided that we allow
causality to go backward in time with backdated state reduction.

Let us start with the similar but simpler case of polarization entanglement.
We consider two photons, labeled $L$\ (left) and $R$ (right), generated at
time $t=0$ in a common location $x_{O}$ and\ in a singlet polarization state.
The spatial and polarization state of the two photons at time $t=0$ is
$\frac{1}{\sqrt{2}}\left\vert x_{O}\right\rangle _{L}\left\vert x_{O}%
\right\rangle _{R}\left(  \left\vert 0\right\rangle _{L}\left\vert
1\right\rangle _{R}-\left\vert 1\right\rangle _{L}\left\vert 0\right\rangle
_{R}\right)  $, where $0$ ($1$) stands for horizontal (vertical) polarization.
At time $t=T>0$, this state has evolved into $\frac{1}{\sqrt{2}}\left\vert
x_{L}\right\rangle _{L}\left\vert x_{R}\right\rangle _{R}\left(  \left\vert
0\right\rangle _{L}\left\vert 1\right\rangle _{R}-\left\vert 1\right\rangle
_{L}\left\vert 0\right\rangle _{R}\right)  $, with the two photons in the two
different locations $x_{L}$\ (on the left) and $x_{R}$ (on the right). If we
measure $\left[  L\right]  $ (the polarization of the left photon)\ at time
$T$ and find the eigenvalue $0$, this implies state reduction on\ $\left\vert
x_{L}\right\rangle _{L}\left\vert x_{R}\right\rangle _{R}\left\vert
0\right\rangle _{L}\left\vert 1\right\rangle _{R}$and that the measurement of
$\left[  R\right]  $, performed (say) at the same time, yields the eigenvalue
$1$. As well known, this logical implication can be backed by the following
causal (deterministic/local) process. We backdate state reduction on
\ $\left\vert x_{L}\right\rangle _{L}\left\vert x_{R}\right\rangle
_{R}\left\vert 0\right\rangle _{L}\left\vert 1\right\rangle _{R}$ to time
$t=0$. Correspondingly $\frac{1}{\sqrt{2}}\left\vert x_{O}\right\rangle
_{L}\left\vert x_{O}\right\rangle _{R}\left(  \left\vert 0\right\rangle
_{L}\left\vert 1\right\rangle _{R}-\left\vert 1\right\rangle _{L}\left\vert
0\right\rangle _{R}\right)  $\ reduces on $\left\vert x_{O}\right\rangle
_{L}\left\vert x_{O}\right\rangle _{R}\left\vert 0\right\rangle _{L}\left\vert
1\right\rangle _{R}$. This can be interpreted as the $L$ photon locally
telling the $R$ photon that its polarization should be $1$, which goes forward
in time back to $\left\vert x_{L}\right\rangle _{L}\left\vert x_{R}%
\right\rangle _{R}\left\vert 0\right\rangle _{L}\left\vert 1\right\rangle
_{R}$, when $\left[  R\right]  $ is measured.

We apply this rationale to quantum computation. To start with, we should break
down the content of register $K$ into content of first qubit and content of
second qubit -- i. e. $\left[  K\right]  $ is broken down into $\left[
K_{0}\right]  $ and $\left[  K_{1}\right]  $. Similarly $\left[  X\right]
$\ is broken down into $\left[  X_{0}\right]  $ and $\left[  X_{1}\right]  $.
$k_{0}$ ($k_{1}$)\ is the eigenvalue obtained by measuring $\left[
K_{0}\right]  $ ($\left[  K_{1}\right]  $) -- indifferently in (\ref{input})
or (\ref{output}). $x_{0}$ ($x_{1}$) is the eigenvalue obtained by measuring
$\left[  X_{0}\right]  $ ($\left[  X_{1}\right]  $)\ -- in (\ref{output}).\ 

We can see that, even allowing causality to go backward in time with backdated
state reduction, it cannot be true that $k_{0},k_{1}$ causes $x_{0}%
=k_{0},x_{1}=k_{1}$. In fact computer science tells us that there is no causal
(deterministic/local) process that goes from cause $k_{0},k_{1}$ to effect
$x_{0}=k_{0},x_{1}=k_{1}$ through one computation of $\delta$; three
computations are required. In other words, it is not true that choosing the
drawer number to hide the ball in on the part of the oracle causes the drawer
number the ball is found in by the algorithm. In this case logical implication
is not backed by a causal process.

For the same reason, reversing the direction of time, it cannot be true that
$x_{0},x_{1}$ causes $x_{0}=k_{0},x_{1}=k_{1}$. In other words, it is not true
that reading the drawer number produced by the algorithm, with no oracle's
choice having been performed as yet, puts the ball in the drawer with that number.

We should look for a different causal process that ends in the effect
$x_{0}=k_{0},x_{1}=k_{1}$\ and involves one computation of $\delta$. To this
end, we note that the implication $(k_{0},k_{1})\rightarrow(x_{0}=k_{0}%
,x_{1}=k_{1})$ is equivalent to $(k_{0},x_{1})\rightarrow(x_{0}=k_{0}%
,x_{1}=k_{1})$, or to $(k_{1},x_{0})\rightarrow(x_{0}=k_{0},x_{1}=k_{1})$.
Correspondingly, we have the two following causal processes:

\begin{itemize}
\item $k_{0},x_{1}$ (the outcomes of measuring $\left[  K_{0}\right]  $ and
$\left[  X_{1}\right]  $)\ causes $x_{0}=k_{0},x_{1}=k_{1}$. Finding, for
example, $k_{0}=0,x_{1}=1$ causes $k_{1}=1,x_{0}=0$ through a single
computation of $\delta$. In fact one bit of the ball location, $k_{0}=0$ (due
to measuring $\left[  K_{0}\right]  $) should be ascribed to the oracle's
choice, the other bit, $x_{1}=1$ (due to measuring $\left[  X_{1}\right]  $)
should be ascribed to the second player -- to her reading at the end of the
algorithm the other bit of the ball location in register $X$, without any
oracle's choice having been performed as yet on the value of that bit. This
other bit is the ball put in that bit. Thus the quantum algorithm has to
search only the bit ascribed to the oracle's choice $k_{0}=0$, which requires
one computation of $\delta$.

\item $k_{1},x_{0}$\ (the outcomes of measuring $\left[  K_{1}\right]  $ and
$\left[  X_{0}\right]  $)\ causes $x_{0}=k_{0},x_{1}=k_{1}$. The discussion is similar.
\end{itemize}

If we measure $\left[  K\right]  $ in state (\ref{input}), or indifferently
backdate to before running the algorithm the outcome of measuring $\left[
K\right]  $ in (\ref{output}), the oracle's choice is pre-fixed before running
the algorithm, say to $k_{0}=0,k_{1}=1$. The second player putting the ball in
$x_{1}=1$, as from the above example, should be replaced by her knowing in
advance one bit of the solution she will find in the future.

Correspondingly, the computation stage of the quantum algorithm can be broken
down as a sum of all the possible histories of a classical algorithm that,
knowing in advance 50\% of the information about the solution, performs the
computations still required to identify the missing information. Each history
is represented in quantum notation as the sequence of two sharp states, one
before and the other after the computation of $\delta$.

For example, in the present case of Grover's algorithm, let us assume that the
second player knows in advance that the oracle's choice is either $k_{0}=0$,
$k_{1}=0$\ or $k_{0}=0$, $k_{1}=1$ (which means knowing in advance 50\% of the
information about the oracle's choice, given that this choice has been
restricted from 4 to 2 possibilities). To establish which is the case, she
should query the black box with either $x_{0}=0$, $x_{1}=0$ or $x_{0}=0$,
$x_{1}=1$. Let us assume it is with $x_{0}=0$, $x_{1}=0$. If the outcome of
the computation is $\delta=1$, this means that $k_{1}=0$. This pinpoints two
possible histories, depending on the initial state of register $V$. History \#
1: initial state $\left\vert 00\right\rangle _{K}\left\vert 00\right\rangle
_{X}\left\vert 0\right\rangle _{V}$, state after the computation $\left\vert
00\right\rangle _{K}\left\vert 00\right\rangle _{X}\left\vert 1\right\rangle
_{V}$. History \#2: initial state $\left\vert 00\right\rangle _{K}\left\vert
00\right\rangle _{X}\left\vert 1\right\rangle _{V}$, state after the
computation $\left\vert 00\right\rangle _{K}\left\vert 00\right\rangle
_{X}\left\vert 0\right\rangle _{V}$. If instead the outcome of the computation
is $\delta=0$, this means that $k_{1}=1$. This pinpoints two other possible
histories. History \# 3: initial state $\left\vert 01\right\rangle
_{K}\left\vert 00\right\rangle _{X}\left\vert 0\right\rangle _{V}$, state
after the computation $\left\vert 01\right\rangle _{K}\left\vert
00\right\rangle _{X}\left\vert 0\right\rangle _{V}$. History \#4 initial state
$\left\vert 01\right\rangle _{K}\left\vert 00\right\rangle _{X}\left\vert
1\right\rangle _{V}$, state after the computation $\left\vert 01\right\rangle
_{K}\left\vert 00\right\rangle _{X}\left\vert 1\right\rangle _{V}$. Etc.

If we sum together all the possible histories, each with a suitable phase
($+1$ or $-1$), and normalize, we obtain the transformation of state
(\ref{input}) into (\ref{secondstage}). This shows that the computation stage
of the quantum algorithm can be broken down into a sum over the histories of a
classical algorithm that knows in advance 50\% of the information about the solution.

This has an important practical consequence: the speed up in terms of number
of oracle's queries comes from comparing two classical algorithms, with and
without advanced information. This allows to characterize the problems liable
of being solved with a quantum speed up in an entirely computer science
framework with no physics involved -- an important simplification -- see
$\left[  1\right]  $.

Furthermore, as we have shown in $\left[  1\right]  $, the history phases that
reconstruct the quantum algorithm are also such that they maximize -- after
the computation of $\delta$ -- the entanglement between registers $K$ and $X$.
Then the final rotation of the basis of register $X$, transforming state
(\ref{secondstage}) into (\ref{output}), transforms entanglement between $K$
and $X$\ into correlation between the outcomes of measuring $\left[  K\right]
$ and $\left[  X\right]  .$ This allows to synthesize the quantum algorithm
out of the advanced information classical algorithm. It is thus a tool for the
search of new quantum speed ups.

We discuss a possible objection to the present explanation of the speed up.
The oracle's choice can be fixed before running the algorithm, say to
$k_{0}=0,k_{1}=1$. In this case the unitary part of the algorithm
(deterministically) produces state (\ref{am}) and quantum measurement of
$\left[  X\right]  $ in (\ref{am}) produces the solution $x_{0}=0$
and$~x_{1}=1$ with probability 1. The objection could be that the quantum
algorithm in this case produces the solution in a deterministic way. Thus the
nondeterministic character of quantum measurement would play no role in the
quantum speed up. We show that this is not the case.

We start without a fixed oracle's choice and represent the data base search
problem (for $N=4$) as the problem of satisfying the nonlinear Boolean network%

\begin{equation}
\delta=AND\left(  y_{0},y_{1}\right)  ,~y_{0}=\sim XOR\left(  k_{0}%
,x_{0}\right)  ,~y_{1}=\sim XOR\left(  k_{1},x_{1}\right)  ,~\delta=1.
\label{boolean}%
\end{equation}
The relation between the Boolean variables $k_{0},~k_{1},~x_{0},$ and $x_{1}$
established by this network is also the relation between the outcomes of
measuring $\left[  K\right]  $ and $\left[  X\right]  $\ in (\ref{output}).
Satisfying this network classically requires trying several computations of
the three gates (discarding those that yield $\delta=0$) -- 2.25 on average.
Instead the quantum algorithm (unitary part and measurement)
nondeterministically satisfies the network with a single computation of the
gates. This produces one of the four possible oracle's choices and the
corresponding solution provided by the second player.

If, in (\ref{boolean}), we fix the values of $k_{0}$ and$~k_{1}$, the
difficulty of the problem remains unaltered. Measuring $\left[  X\right]
$\ in (\ref{am})\ still nondeterministically satisfies a nonlinear Boolean
network (with the values of $k_{0}$\ and $k_{1}$ pre-fixed and the values of
$x_{0}$\ and $x_{1}$\ unknown), the only difference is that the result is
definite (produced with probability 1), but this is so just because this
network admits only one solution. We should not mistake the nondeterministic
production of a definite outcome for a deterministic production.

In conclusion, we believe that, in quantum algorithms, heuristics went ahead
of theory, and that the present explanation of the quantum speed up provides a
useful theoretical clarification.

\textbf{Acknowledgements}

The author thanks for useful discussions: Vint Cerf, Artur Ekert, David
Finkelstein, Hartmut Neven, Daniel Sheehan, and Henry Stapp.

\textbf{Bibliography}

$1.$ Castagnoli, G.: Quantum algorithms know in advance 50\% of the solution
they will find in the future. http://arxiv.org/pdf/0906.1811 and
http://www.springerlink.com/
openurl.asp?genre=article\&id=doi:10.1007/s10773-009-0143-6 (2009)

$2.$\ Castagnoli, G.: The 50\% advanced information rule of the quantum
algorithms. Int. J. Theor. Phys. vol. 48, issue 8, 2412 (2009)

$3.$\ Grover, L. K.: A fast quantum mechanical algorithm for data base search.
Proc. 28th Ann. ACM Symp. Theory of Computing (1996)

\end{document}